\documentstyle[pre,eqsecnum,aps]{revtex}
\begin{document}
\draft
\date{\today}
\title{PHASE ORDERING KINETICS OF \\
ONE-DIMENSIONAL NON-CONSERVED SCALAR SYSTEMS}
\author{A. D. Rutenberg and A. J. Bray\cite{byline}}
\address{Theoretical Physics Group \\
Department of Physics and Astronomy \\
The University of Manchester, M13 9PL, UK \\
cond-mat/9404046  }
\maketitle
\begin{abstract}
We consider the phase-ordering kinetics of one-dimensional scalar systems.
For attractive long-range ($r^{-(1+\sigma)}$)
interactions with $\sigma>0$, ``Energy-Scaling''
arguments predict a growth-law of the average domain size
$L \sim t^{1/(1+\sigma)}$ for all $\sigma >0$.
Numerical results for $\sigma=0.5$, $1.0$, and $1.5$ demonstrate both
scaling and the predicted growth laws.
For purely short-range interactions, an approach of Nagai and Kawasaki
is asymptotically exact. For this case, the equal-time correlations
scale, but the time-derivative correlations break scaling. The short-range
solution also applies to systems with long-range interactions when
$\sigma \rightarrow \infty$, and in that limit the
amplitude of the growth law is exactly calculated.
\end{abstract}

\section{Introduction}
\label{SEC:INTRO}

The quench of a system from a disordered phase to an ordered phase
is a non-equilibrium process in which
energy is dissipated and any topological defects are eliminated.
Typically, the system develops a scaling structure, with a single
time-dependent
length scale, as the growing broken-symmetry
phases compete to select the ordered phase \cite{Gunton83}.
In this paper we discuss non-conserved
one-dimensional (1D) scalar systems, where the competing
phases are domains of uniform magnetization and the topological defects
are domain walls. We limit our discussion to the late stages of phase ordering,
when the domains are much larger than the interface width and the
scaling structure, if any, has been established.
The phase-ordering dynamics is driven by interactions between domains, rather
than by the curvature of domain walls which drives the dynamics of short-range
systems in higher dimensions \cite{Allen79}.
For short-range 1D systems this interaction is through the exponential
tail of the domain-wall profile \cite{Kawasaki82}.
For long-range systems, the domains have a direct $r^{-(1+\sigma)}$
interaction.

Following work in dimensions greater than one
\cite{Bray93,Hayakawa93a,Hayakawa93b}, there
has been recent interest in 1D systems with attractive long-range interactions.
This paper follows up a broader treatment that
determined growth laws for for a wide variety of scaling systems, including
1D systems with $0 <\sigma <1$ \cite{Bray94}.
Lee and Cardy \cite{Lee93} found analytic and numerical indications of
scaling violations and anomalous growth laws for $0<\sigma \leq 1$,
in addition to results consistent with scaling for $\sigma >1$.
This motivated us to extend their numerical studies for systems
with $\sigma=0.5$, $1.0$, and $1.5$, and to extend our theoretical
treatment to all $\sigma>0$ (section \ref{SEC:LR}).
We find that larger systems than those studied by Lee and Cardy
exhibit both scaling and our predicted
growth laws (section \ref{SEC:SIMULATION}).

Short-range 1D scalar systems have been
studied both numerically \cite{Nagai83} and experimentally
\cite{Ikeda83}.  Nagai and Kawasaki have presented a solution
assuming uncorrelated domain sizes
\cite{Nagai86,Kawasaki88}.
We rederive their solution with a simplified approach,
and show that it is asymptotically exact for systems
with purely short-range interactions, and also
for systems with long-range interactions in the limit
$\sigma \rightarrow \infty$
(section \ref{SEC:SHORTRANGE}).

A generic energy functional for a 1D system
with a scalar order parameter, $\phi(x)$, and short-range interactions, is
\begin{equation}
\label{EQN:HAMILTONIAN}
H[\phi] = \int d x\,\left[ (d\phi/dx)^2 + V(\phi) \right]\ ,
\end{equation}
where $V(\phi)$ is a double well potential such as
\begin{equation}
\label{EQN:POTENTIAL}
V(\phi)= (\phi^2 -1)^2.
\end{equation}
We restrict our attention to a potential
with symmetric quadratic minima \cite{soft}.
We can also add long-range attractive interactions,
\begin{eqnarray}
\label{EQN:INTERACTION}
        H_{\text{LR}} &=& \int dx \int dr
        \frac{\left[ \phi(x+r)-\phi(x) \right]^2}
                { r^{1+\sigma}},     \nonumber \\
        & \simeq & A_{\sigma} L_\infty \int dk k^\sigma \phi_k \phi_{-k},
\end{eqnarray}
where $L_\infty$ is the system size and we take
$\sigma >0$ for a well defined thermodynamic limit.  The long-range
interactions dominate the 1D dynamics when they exist, so we
have ignored the short-range component of (\ref{EQN:INTERACTION}) for
$\sigma \geq 2$.

After a temperature quench into the ordered phase, the
equation of motion for the ordering kinetics for non-conserved systems
with purely dissipative dynamics is \cite{Hohenberg77}:
\begin{equation}
\label{EQN:DYNAMICS}
\partial_t {\phi} = - \delta H/\delta {\phi}.
\end{equation}
For short-range systems, and long-range systems with $\sigma>1$ \cite{Dyson69},
there is an ordered phase only at $T=0$.
We restrict ourselves to $T=0$, with no thermal noise, for {\em all} cases,
since temperature is an ``irrelevant variable''
within an ordered phase \cite{Bray89}.

\section{Long-range attractive interactions}
\label{SEC:LR}

We first treat systems with long-range attractive
interactions of the form (\ref{EQN:INTERACTION}).
As was shown by Lee and Cardy for Ising spins \cite{Lee93},
the dynamics of the system is captured by the motion of the domain walls.
Consider two sharp domain walls of the same sign at $r_i$ and $r_j$, as shown
in
figure \ref{FIG:WALLINT}.
Their interaction energy, $E_{ij}$, is the part of
$H_{\text{LR}}$ that depends
on their separation:
\begin{eqnarray}
\label{EQN:WALLINT}
     E_{ij}  &=&  \left\{
        \begin{array}{c}
         -(1-\sigma)^{-1} |r_j-r_i|^{1-\sigma}, \ \ \ \ \ \  \sigma \neq 1, \\
         -\ln{|r_j-r_i|} , \ \ \ \ \ \ \ \ \ \ \ \ \ \sigma=1,
         \end{array}
        \right.
\end{eqnarray}
where we have scaled out a factor of $8/\sigma$.
Domains of opposite sign introduce an overall $-1$ factor
to the energy.
This interaction energy leads to a force,
\begin{eqnarray}
\label{EQN:FORCE}
        F_{ij}  &=& -\partial E_{ij}/\partial r_j,  \nonumber \\
                &=& \pm l^{-\sigma},
\end{eqnarray}
where $l=|r_j-r_i|$ and the force is repulsive between walls of the same
sign and attractive between walls of the opposite sign.
The domain wall velocity is proportional to the summed pairwise
forces acting on it \cite{diss}:
\begin{equation}
\label{EQN:UPDATE}
        \dot{r_i} =  {\sum_j}' F_{ij}.
\end{equation}
When adjacent domain walls meet, they annihilate --- this drives the
non-equilibrium coarsening process.  We use the domain-wall
dynamics described by equations
(\ref{EQN:FORCE}) and (\ref{EQN:UPDATE}) unless specifically noted.

This domain-wall dynamics also holds with the soft potential
$V(\phi)$ in (\ref{EQN:POTENTIAL}). Fixing the domain wall positions
$\{ r_i \}$, we use the quasi-static solution of
(\ref{EQN:DYNAMICS}) expressed in real space:
\begin{eqnarray}
\label{EQN:STATIC}
        0 &\approx& \delta H /\delta \phi, \nonumber \\
        & =&    \partial_x^2 \phi-4 \phi (\phi^2-1)+\int_{-\infty}^\infty dx'
        \left[ \phi(x')-\phi(x) \right]/|x'-x|^{1+\sigma}, \nonumber \\
        & \simeq & \phi(x) \left\{
        -4 \left[ \phi(x)^2-1 \right] -2 \sum_{n=0}^\infty
                \left[
\int_{r_{i-2n-1}}^{r_{i-2n}} dx'/|x'-x|^{1+\sigma}
+\int_{r_{i+2n+1}}^{r_{i+2n+2}} dx'/|x'-x|^{1+\sigma} \right] \right\},
\end{eqnarray}
where $x \in (r_i,r_{i+1})$.
We use the saturated values
$\phi \approx \pm 1$ within the integral,
so that the integrals are restricted to phases with $\phi(x')=-\phi(x)$.
We also assume that $x$ is far from all domain walls, so that
we neglect $\partial_x^2 \phi$.
This leads to $\phi(x) = \pm (1- p(x))$ with
\begin{equation}
\label{EQN:PROFILE}
p(x) \simeq \frac{1}{4\sigma}  {\sum_{m=0}^\infty} (-1)^m
 \{ |x-r_{i-m}|^{-\sigma} + |x-r_{i+1+m}|^{-\sigma} \}
\end{equation}
at large distances from other domain walls (compare \cite{Hayakawa93b}).
The force corrections due to the asymptotic profile can be easily estimated
by the energy per unit length of adding to the ends of the power-law profile
between two defects as they move apart. From the potential term in
(\ref{EQN:HAMILTONIAN}), $V(\phi)= (\phi^2-1)^2 \sim l^{-2\sigma}$,
where the defects are a distance $l$ apart.
{}From the gradient squared-term, the effective force is $l^{-2-2\sigma}$,
which
at late times is always much less than from the local potential.
Since we are assuming a quasi-static configuration, the
change in energy due to the long-range interactions is of the
order of that due to the potential term, so is also $l^{-2\sigma}$.
These corrections are negligible in comparison to
the $l^{-\sigma}$ interactions of ``sharp'' domain walls
(\ref{EQN:FORCE}), which we use without loss of generality.

\subsection{``Energy-scaling'' theory for growth laws}
\label{SEC:ENSCALE}

We predict the growth-law of the characteristic length, $L(t)$, for scaling
systems by considering the energetics of the system \cite{Bray94}.
The scaling behavior, or dependence on the length-scale, of
the energy density is described by the long-range
interactions (\ref{EQN:INTERACTION}):
\begin{equation}
\label{EQN:LONGRANGE}
\epsilon \sim \int dk k^\sigma\,
\left< {\phi}_{ k} {\phi}_{-{ k}} \right>,
\end{equation}
where the angle brackets indicate an average over initial conditions.
We independently calculate the rate of energy-density dissipation
by integrating the dissipation in each Fourier mode and using the dynamics
(\ref{EQN:DYNAMICS}) expressed in momentum
space:
\begin{eqnarray}
\label{EQN:ENDISS}
d{\epsilon}/dt &=& \int dk \left< (\delta H/\delta {\phi}_{k})
        \partial_t {\phi}_{k} \right>,  \nonumber \\
&=& - \int dk \,
        \left< \partial_t {\phi}_{k}
        \partial_t {\phi}_{-k}
        \right>\ .
\end{eqnarray}
We use the equal-time correlation function,
$S(k,t) \equiv  \left< {\phi}_{k} {\phi}_{-{k}} \right>$,
to determine the energy density, and the
time-derivative correlation function,
$T(k,t) \equiv \left< \partial_t {\phi}_{k}
\partial_t {\phi}_{-k}  \right>$,  to
determine the rate of energy-density dissipation.

If we assume scaling --- that the correlations are described by one
time-dependent length-scale $L(t)$ but not by any microscopic scales
--- then dimensional analysis
determines the structure function:
\begin{equation}
\label{EQN:STRUCT}
S({k},t) = L(t)\,g(kL(t)),
\end{equation}
where the scaling function $g(x)$ has no time-dependence.
This is also known as the dynamic scaling hypothesis \cite{Furukawa85}.
We apply the same scaling hypothesis to the two-time correlation
function: $S(k,t,t') \equiv \left< {\phi}_{k} (t) {\phi}_{-{k}} (t')\right>
= k^{-1} \tilde{g}(kL(t),kL(t'),t/t')$. This leads to a scaling form
for the time-derivative correlation function:
\begin{eqnarray}
        \left< \partial_t {\phi}_{k} \partial_t
        {\phi}_{-{k}} \right> &=& \partial_t \partial_{t'}|_{t=t'}
                S(k,t,t'), \nonumber \\
\label{EQN:FULLTTSCALE}
        &=& (\dot{L}/L)^2 L \tilde{g}_{xy}(kL)+t^{-2} L \tilde{g}_z(kL),  \\
\label{EQN:TTSCALE}
        &\simeq& (\dot{L}/L)^2 L g_{\text{tt}}(kL),
\end{eqnarray}
where for power-law growth, with or without logarithmic factors,
the $\tilde{g}_z$ contribution does not dominate at late times.

If the energy
integrals (\ref{EQN:LONGRANGE}) and (\ref{EQN:ENDISS}) are convergent
then we use the scaling forms and change variables to get the $L$ and
$\dot{L}$ dependence.  Certainly, the integrals are convergent for $kL \ll 1$
if the thermodynamic limit exists.  However, the integrals may diverge
for $kL \gg 1$. For $S(k,t)$, that limit is given by Porod's law
\cite{Porod82},
in which the structure at $k L \gg 1$ is proportional to the wall
density $\rho_{\text{wall}} \sim L^{-1}$,
\begin{equation}
\label{EQN:POROD}
S({k},t) \sim L^{-1}\,k^{-2}\ ,\ \ \ \  kL \gg 1.
\end{equation}
With this and the scaling form (\ref{EQN:STRUCT}) we determine the energy
density (\ref{EQN:LONGRANGE}):
\begin{equation}
\label{EQN:ELR}
\epsilon  \sim  \left\{
        \begin{array}{c}
        L^{-1}\,\xi^{1-\sigma}\ ,\ \ \ \ \ \ \ \ \ \ \ \ \ \sigma>1,  \\
        L^{-1}\,\ln{(L/\xi)}\ ,\ \ \ \ \ \ \ \ \ \sigma=1, \\
        L^{-\sigma}\ ,\ \ \ \ \ \ \ \ \ \ \ \ \ \ \ \ \ \ \ \  \sigma<1.
        \end{array}
        \right.
\end{equation}
For $\sigma>1$ the integral diverges for $kL \gg 1$ and
the energy density is proportional to the domain-wall density,
$\rho_{\text{wall}} \sim L^{-1}$;
for $\sigma=1$ the logarithmic divergence of the integral indicates that
structure on all scales contributes to the energy-density; while
for $\sigma < 1$ the integral converges and the
energy is dominated by the long-range interaction
between domains.

The {\em interaction} energy-density,
$\epsilon_{\text{int}}$, of domains at the characteristic
scale is the convergent $kL \approx 1$
part of the energy integral. This energy-density of interaction is what
drives the motion of well separated domain walls:
\begin{equation}
\label{EQN:EINT}
\epsilon_{\text{int}} \sim L^{-\sigma}.
\end{equation}
To determine the growth law,
we only need the time-derivative of $\epsilon_{\text{int}}$, which comes
from domain walls moving at the characteristic velocity $\dot{L}$.
Near a particular domain wall the order parameter comoves with the domain wall:
\begin{equation}
\label{EQN:COMOVING}
\partial_t{{\phi}}= - v \partial\phi / \partial x.
\end{equation}
The contribution of these slowly moving domain walls in momentum space is
\begin{eqnarray}
\label{EQN:TTDOTL}
\left< \partial_t \phi_{ k}
        \partial_t {\phi}_{ -k}\right>_{\text{int}} &\sim&
                \dot{L}^2 k^2 \left< {\phi}_{k} {\phi}_{ -k}\right>, \nonumber
        \\
        &\sim& \dot{L}^2/L, \ \ \ \ \ \ \ \ \ \ \ \ \ \  kL \gg 1,
\end{eqnarray}
where we use Porod's law to obtain the second line.  While this expression
satisfies the scaling form (\ref{EQN:TTSCALE}), it is {\em not}
the full time-derivative correlation function, merely the part that contributes
to the dissipation of the interaction energy $\epsilon_{\text{int}}$.
We use this contribution in (\ref{EQN:ENDISS}) to find
\begin{equation}
\label{EQN:RHS}
d{\epsilon}_{\text{int}}/dt
 \sim   \xi^{-1}\,\dot{L}^2/ L .
\end{equation}
The integral diverges for $kL \gg 1$, but the divergence reflects
the internal structure of the domain walls,
rather than small separations with respect to $L(t)$.
We compare the rate of energy-density dissipation to the time derivative
of (\ref{EQN:EINT}) to obtain
$\dot{L} \sim L^{-\sigma}$, which determines the growth law:
\begin{equation}
\label{EQN:GROWTH}
 L(t) \sim t^{1/(1+\sigma)},
\end{equation}
for all $\sigma >0$.

We check the consistency of this approach by calculating the full
$d{\epsilon}/dt$. This is
easily done in real space, where from (\ref{EQN:ENDISS})
we have
\begin{eqnarray}
\label{EQN:REALENDISS}
d{\epsilon}/dt
        &=& -  L_\infty^{-1} \int dx
        \left< \partial_t {\phi}(x) \partial_t {\phi}(x) \right>\ ,
                        \nonumber \\
        &\sim& L^{-1} \left< v^2 \right>,
\end{eqnarray}
where we have used (\ref{EQN:COMOVING}) to obtain the second line.
The result is the density of domain walls, $\rho_{\text{wall}} \sim L^{-1}$,
times the average square-velocity.
Either $\left< v^2 \right> = \dot{L}^2$ and the energy-density is the
same as $\epsilon_{\text{int}}$, or the energy is dominated by the domain wall
energy and the energy-dissipation is
dominated by rapidly annihilating domain walls.

For domain sizes
$l \ll L$, the force due to nearest-neighbors will set the scale of the
net force on a domain wall, so from (\ref{EQN:FORCE}):
\begin{equation}
\label{EQN:VEL}
|v(l)| \approx \beta l^{-\sigma}, \ \ \ \ \ \ \ \ \ l \ll L,
\end{equation}
where the constant prefactor $\beta$, where $0<\beta \leq 1$,
proves sufficient to parameterize any shielding due to other domain walls.
If $n_t(l) $ is the number distribution of domain sizes,
so that $n_t(l) \delta l$ is the number of domains of sizes in the
interval $[l, l+\delta l]$, then the number flux of
collapsing domains is given by
$j(l)= n_t(l) \left[ - 2|v(l)| \right]$ for $l \ll L$.
The factor of two follows since the
collapse rate of a small domain is twice the speed of each of its walls.
When a domain collapses its domain walls vanish, merging
the two adjacent domains. Hence, every
domain that collapses decreases the total number of domains by $2$.
Matching $2j(\xi)$
to the rate of change of domain number gives
$2 j(\xi,t) = \partial_t N = - \dot{L} L_\infty /L^2$,
where $N = L_\infty / L = \int_\xi^\infty dl n_t(l)$.
Since domain walls annihilate at a constant
rate over timescales much less than $L/\dot{L}$, the domain wall flux is
constant for scales $l \ll L$: $j(l) \approx j(\xi)$. This leads to
\begin{equation}
\label{EQN:COLLAPSE}
 n_t(l) \approx \frac{N}{4 \beta L} \dot{L} l^{\sigma}, \ \ \ \ \ \ l \ll L,
\end{equation}
where $\beta = 1$ if no shielding of small collapsing domains occurs.

With (\ref{EQN:VEL}) and (\ref{EQN:COLLAPSE})
we determine the contribution of domains of scale $l \ll L$
to the average square velocity:
\begin{eqnarray}
\label{EQN:VELSQUARE}
\left< v^2 \right>_{small} &=& \frac{\int_\xi^L dl v^2(l) n_t(l) }
                {\int_\xi^L dl n_t(l)}, \nonumber \\
         &\sim & \dot{L} L^{-1} \int_\xi^L dl\, l^{-\sigma} , \nonumber \\
        &\sim& \left\{
                \begin{array}{c}
        \xi^{1-\sigma} \dot{L} L^{-1}, \ \ \ \ \ \ \ \ \ \ \ \ \sigma >1, \\
        \dot{L} L^{-1} \ln{(L/\xi)}, \ \ \ \ \ \ \ \ \sigma=1,  \\
        \dot{L} L^{-\sigma}, \ \ \ \ \ \ \ \ \ \ \ \ \ \ \ \sigma<1.
                \end{array}
                \right.
\end{eqnarray}
For $\sigma<1$ small domains do not dominate the
average and, since $\dot{L} \sim L^{-\sigma}$, we have
$\left< v^2 \right> \sim \dot{L}^2$.
Indeed, for this case $\epsilon \sim \epsilon_{\text{int}}$, and the
energetics is controlled by domains at the characteristic scale.
For $\sigma \geq 1$, the energy-density
dissipation, calculated with (\ref{EQN:REALENDISS}) and (\ref{EQN:VELSQUARE}),
is consistent with (\ref{EQN:ELR}). However, the energy-density
dissipation is dominated by small annihilating domains and does not determine
the growth law of the average domain size, for which we have used
$\epsilon_{\text{int}}$.

We have numerically studied the average square velocity of domain walls,
as well as the average velocity. In Figure
\ref{FIG:VEL} we show $\left<v^2 \right>/\dot{L}^2$ and
$\left<|v| \right>/\dot{L}$ vs. time.
The scaled velocity is approximately constant
away from early time transients and late time small number effects.
Indeed, a similar analysis to that above shows
$\left< |v| \right> \sim \dot{L}$ for all $\sigma>0$.
The average
square-velocity scales well for $\sigma=0.5$. However for $\sigma=1.0$ and
$1.5$, $\left<v^2 \right>$ is much larger
than $\dot{L}^2$ --- corresponding to the
dominance of small rapidly annihilating domains for $\sigma \geq 1$.
While sparse statistics for extremely small domains
leads to sporadic undercounting of $\left< v^2 \right>$ for $\sigma \geq 1$,
the scaled square-velocity for $\sigma=1.5$ agrees with
$\left<v^2 \right>/\dot{L}^2 \sim (L \dot{L})^{-1} \sim t^{1/3} $, which is
shown by the straight line.

\subsection{Simulations}
\label{SEC:SIMULATION}

To check these growth laws and the scaling of the correlation functions, we
perform computer simulations for $\sigma=0.5$, $1$, and $1.5$.
These are the cases considered by Lee and Cardy
\cite{Lee93}, and our treatment is motivated by theirs.
We simulate the evolution of systems with sharp domain
walls of alternating sign
which annihilate when they meet. These domain walls are treated as
particles. They interact with long-range
forces, $f(l) = \pm l^{-\sigma}$, where $l$ is the domain wall separation.
The force is repulsive or attractive for like or unlike domain
walls, respectively.

We first measure the number of domain walls $N(t)$
as a function of time, starting
from a  fixed number placed randomly \cite{size}, and continuing
until none remain. With a domain wall velocity proportional to the net
force (\ref{EQN:UPDATE}), we use a simple Euler update for each domain wall
location, $\Delta r_i = \dot{r}_i \Delta
t$. Only the early-time behavior depends on the size and nature
of the timestep $\Delta t$, if $\Delta t$ is small enough.  We use a timestep
proportional to the time ($\Delta t = t/20$), which for power-law growth
results in a fixed minimum {\em scaled} domain size that annihilates
in one timestep \cite{timestep},
\begin{equation}
\label{EQN:LMIN}
l_{min}/L \propto (\Delta t/t)^{1/(1+\sigma)}.
\end{equation}
The choice of boundary condition only affects the late-time results.
We use free boundary conditions \cite{free}
for which small number effects are not seen until the average number
of domain walls $N \lesssim 1$.  We average over at least a thousand
independent runs for most systems, and over at least 50 runs for the
largest system size with 12800 initial domain walls.

We expect the average domain size $L(t)$, which is the inverse of the
domain wall density $\rho(t)$, to have our predicted growth law $L(t) =
t^{1/(1+\sigma)}/\Phi^\ast(\sigma)$. The ratio of the measured domain wall
density to the predicted, $\rho(t) t^{1/(1+\sigma)}$ up to the amplitude
factor,  should be a scaling function of dimensionless combinations of lengths:
\begin{equation}
\label{EQN:DENSITYSCALING}
\rho(t) t^{1/(1+\sigma)} = \Phi \left[ \rho(t)/\rho_0, 1/(\rho(t) L_\infty)
\right].
\end{equation}
In Figure \ref{FIG:ALLGROWTH} we plot, for various $\sigma$,
the scaling function $\Phi$ as a function of the number of domain walls, $N$,
on the left and of $\rho(t)/\rho_0$ on the right.
We see an early-time transient (i.e. $\rho_0/\rho(t)$ of order $1$)
which is independent of the system size, and a small number regime
(i.e. $ N = \rho(t) L_\infty$ small) at late times
which is independent of the initial density \cite{size}.
At intermediate times the scaling function is constant,
$\Phi=\Phi^\ast(\sigma)=\Phi[0,0]$,
independent of both the initial density and the system size.
This indicates a scaling regime, growing with system size, that confirms
our predicted growth law $L(t) = t^{1/(1+\sigma)}/\Phi^\ast(\sigma)$.
In table \ref{TAB:G} we plot the amplitude $\Phi^\ast(\sigma)$,
and also the fixed point amplitude used by Lee and Cardy
$g_R^\ast(\sigma)=\Phi^\ast(\sigma) \left[ 2(1+\sigma) \right]^{1/(1+\sigma)}$
 \cite{Lee93}. We include the exact amplitude for $\sigma \rightarrow \infty$
that we derive in section \ref{SEC:SHORTRANGE} (see equation \ref{EQN:PHI}).

In Figure \ref{FIG:BOUNDGROWTH}, we
plot $\Phi$ vs. $N$  with two different
boundary conditions and $\sigma=0.5$. The scaling regime
is of different size in the two systems, but
the scaling value $\Phi^\ast(\sigma)$ is unchanged.
The difference is only in the small number regime ($N \lesssim 20$).
We note that systems with periodic boundary conditions and less than
800 initial domain walls, as studied by Lee and Cardy for $\sigma=0.5$
\cite{Lee93}, do not probe the scaling regime.  We find the expected
scaling growth law, $L \sim t^{2/3}$ for $\sigma=0.5$, in contrast to their
numerical results.

The ratio of the measured energy-density,
$\epsilon(t)$, to the predicted, $\epsilon$
(\ref{EQN:ELR}), should be a new scaling function of dimensionless combinations
of variables:
\begin{equation}
\label{EQN:DENSITYENERGY}
\epsilon(t)/\epsilon = \Phi_E \left[ \rho(t)/\rho_0, 1/(\rho(t) L_\infty)
\right].
\end{equation}
In the scaling regime we expect $\Phi_E=\Phi_E[0,0]$
to be constant. We
plot $\epsilon(t)/\epsilon$ vs. $\rho(t)/\rho_0$
in Fig \ref{FIG:ENERGY} for $\sigma=0.5$ and $1.0$,
where we calculate the predicted $\epsilon$, up to a constant factor,
from the measured length-scale $L(t)= L_\infty/N$.
We use periodic boundary conditions to avoid strong end-effects in the energy.
The domain walls interact with an
infinite-ranged force law so that {\em all} images of a domain wall around the
periodic system are seen (see Appendix \ref{APP:FORCELAW}).
Note that only $21$ runs are done for the largest $\sigma=1.0$ system.
We see the expected early-time
transient independent of system size, an intermediate scaling regime
growing with system size, and a small-number late-time regime.
The energy density for $\sigma=1.5$ is dominated by the core
energy and is simply proportional to the number of domain walls.

While the growth laws agree with the scaling predictions, to investigate
scaling we must consider the scaling of the correlation functions directly.
If scaling holds, the ``Energy-Scaling'' argument
determines the growth laws, but the growth laws are not sufficient
to determine scaling.  To study correlations, we use periodic
boundary conditions with infinite ranged forces. We also use
initial conditions randomly chosen from the $\sigma \rightarrow \infty$
fixed-point distribution (see equation \ref{EQN:FIXEDDIST}), which reduces
the initial transients for all $\sigma$ investigated.
We generally use $\Delta t = t/20$, and average over at least 10000
independent runs, but we present some data at small scaled distances with
a finer timestep, $\Delta t= t/80$, where we average over 1000 runs.

We first plot the scaled number distribution
of domain sizes, $F_L(l/L)=L n_t(l)/ N$,
against the scaled domain size, $x=l/L$, in
Figure \ref{FIG:DOMSIZE}. The exact result for $\sigma \rightarrow \infty$
is also shown. The straight lines are the fit, for small $x$, to the
power-law behavior from (\ref{EQN:COLLAPSE}). The deviation from the power law
at very small $x$ is a result of the geometrically increasing time-step
in the simulations. Domains of size below $l_{\text{min}}$ (\ref{EQN:LMIN})
tend to be randomly distributed.
The filled points have four times as many updates
per timescale, and are correspondingly closer to the power-law.
We tabulate the fitted values of $\beta$ in table \ref{TAB:BETA}. The
decreasing values of $\beta$ for decreasing $\sigma$
indicate stronger shielding and greater correlations between
domains.

To obtain the correlation functions,
we relate the gradient of the field to the positions of the sharp domain walls:
\begin{equation}
\label{EQN:DELPHI}
\partial_x \phi (r)= 2 \sum_{i=1}^{N} \delta(r-r_i)(-1)^i,
\end{equation}
where the sum is over all domain walls, of alternating sign, at positions
$\{ r_i \}$.  This equation gives the correct discontinuity of the
field, $2 (-1)^i$, at the domain walls.
Similarly, the time derivative of the field is
\begin{equation}
\label{EQN:PHIT}
\partial_t \phi (r)=-2\sum_{i=1}^N \delta(r-r_i) (-1)^i v_i,
\end{equation}
where $v_i$ is the signed velocity of the domain wall.
We then easily calculate
\begin{eqnarray}
\label{EQN:2CORR}
 C''(r,t) &=&  - \left< \partial_r\phi(r+r_0)
                \partial_{r_0} \phi(r_0)\right>_{r_0}, \nonumber \\
        &=& - 4 L_\infty^{-1}
        \int dr_0 \sum_{ij} (-1)^{i+j} \delta(r+r_0-r_i) \delta(r_0-r_j),
         \nonumber \\
        &=& L^{-2} \left[ - 4 \delta(x) - 2 N^{-1} \sum_{i \neq j} (-1)^{i+j}
                \left\{ \delta(x-|x_{ij}|)+ \delta(x+|x_{ij}|) \right\}
\right],
\end{eqnarray}
and
\begin{eqnarray}
\label{EQN:2TCORR}
 T(r,t) &=&
\left< \partial_t{\phi}(r+r_0) \partial_t{\phi}(r_0)\right>_{r_0}, \nonumber \\
        &=&  4 L_\infty^{-1}    \int dr_0 \sum_{ij} (-1)^{i+j} v_i v_j
                \delta(r+r_0-r_i) \delta(r_0-r_j), \nonumber \\
        &=& (\dot{L}/L)^2 \left[ 4 \left<v^2 \right> \dot{L}^{-2} \delta(x)+
                2 N^{-1} \sum_{i \neq j} (-1)^{i+j}
                v_i v_j \dot{L}^{-2}
                \left\{ \delta(x-|x_{ij}|)+ \delta(x+|x_{ij}|) \right\}
\right],
\end{eqnarray}
where $x= r/L$, $x_{ij}= (r_j-r_i)/L$, $L_\infty = N L(t)$, and
$\left< v^2 \right>$ is the average square velocity of a domain wall.
The real-space scaling forms are
\begin{equation}
C''(r,t)= L^{-2}f''(r/L(t)),
\end{equation}
\begin{equation}
T(r,t)= (\dot{L}/L)^2 g(r/L),
\end{equation}
where the scaling functions $f''(x)$ and $g(x)$ are
time-independent if scaling holds.
We plot $f''(x)$ in the top row and $g(x)$ in the bottom row
of figure \ref{FIG:ALLCORR} for
$\sigma=0.5$, $1.0$, and $1.5$. Shown are two times differing by a factor
of $4$, both chosen from the scaling region of the
scaling function $\Phi$ (\ref{EQN:DENSITYSCALING}). There
are no discernable scaling violations.
The solid curves through the data are parameterless estimates
assuming independently collapsing domains, which should be valid for small
$x>0$:
\begin{eqnarray}
\label{EQN:ESTF}
        f''(x) &\approx& 2 N^{-1} \sum_{<ij>} \delta(x-|x_{ij}|), \nonumber \\
                &=& 4 F_L(x),
\end{eqnarray}
 and
\begin{eqnarray}
\label{EQN:ESTG}
        g(x) &\approx&  2 N^{-1} \sum_{<ij>} \left[ v_{ij}/\dot{L} \right]^2
                \delta(x-|x_{ij}|),     \nonumber \\
             &\approx& 4 \left[ v(x L)/\dot{L} \right] ^2 F_L(x), \nonumber \\
                &\approx& 4 \beta^2
                \left[ \Phi^\ast(\sigma) \right]^{2(1+\sigma)}
                                        (1+\sigma)^2 x^{-2 \sigma} F_L(x),
        \ \ \ \ \ \ \ \ x \ll 1,
\end{eqnarray}
where the sum is over neighboring domain walls,
$\Phi^\ast(\sigma)$ is measured from figure \ref{FIG:ALLGROWTH}, and
$F_L(x)$ is taken from figure \ref{FIG:DOMSIZE}.
Following equation (\ref{EQN:VEL}),
we have used $v(l)=\beta l^{-\sigma}$ with $\beta$ from table \ref{TAB:BETA}.
We have also used the scaling amplitude of the growth law,
$L(t)=t^{1/(1+\sigma)}/\Phi^\ast(\sigma)$.
For $f''(x)$, the estimate (\ref{EQN:ESTF}),
assuming independently collapsing domains,
is excellent at small $x$.  For $g(x)$, the estimate
(\ref{EQN:ESTG}) agrees at very small $x$.
An intermediate regime, which grows with decreasing $\sigma$,
deviates from the estimate and indicates growing correlations
between domain wall velocities in different domains.
Note that due to the geometrically increasing timestep (\ref{EQN:LMIN})
in our simulations, the domain size distribution deviates from the expected
$F_L(x) \sim x^\sigma$ for $x \ll 1$ (see figure \ref{FIG:DOMSIZE}).
For $\Delta t/t \rightarrow 0$, $f''(x)$ and $g(x)$ should follow
the power-laws indicated by the offset straight lines in the
same regime where the estimates
(\ref{EQN:ESTF}) and (\ref{EQN:ESTG}) hold. This is evident for $f''(x)$
with the data for $\Delta t/t =1/80$, but a fixed time-step is needed
to check the indicated power-law regime in $g(x)$.

We have not shown the $4 \left< v^2 \right> \dot{L}^{-2} \delta(x)$
contribution to $g(x)$. From equation (\ref{EQN:VELSQUARE}),
we see that this {\em breaks} scaling for
$\sigma \geq 1$ due to small rapidly annihilating domains
(see also figure \ref{FIG:VEL}).  Hence, $T(k,t)$ has a
constant contribution that breaks
scaling for {\em all} $k$ for $\sigma \geq 1$.
However, this does not change our ``Energy-scaling'' argument, which depends,
through $\epsilon_{\text{int}}$, only on domains of the characteristic size.

\section{Short-range interactions: continuous spin model}
\label{SEC:SHORTRANGE}

For systems with only short-range interactions, the domain wall profiles
lead to interactions between adjacent domain walls
which drive the phase-ordering. From the dynamical equation
(\ref{EQN:DYNAMICS}),  the equilibrium profile of
an isolated domain wall is determined
by $d^2 \phi/dx^2 = 4 \phi ( \phi^2 -1)$ ---
which gives an exponential tail to the domain wall profile far from
the domain wall. The domain walls, treated as particles,
are attracted  by a corresponding exponential interaction
$V(r) = V_0 e^{-|r|/\xi}$, where $\xi$ is the effective width of the
domain wall \cite{Kawasaki82}, and a resulting pairwise force
$F(r) = -dV/dr = V_0/\xi e^{-|r|/\xi}$.
The domain wall velocities are then simply proportional to their net force.
This case has been studied numerically \cite{Nagai83},
and has been solved by Nagai and Kawasaki \cite{Nagai86} with the
assumption of no correlations between domain sizes
in the fixed point distribution.
We rederive their results using a simpler approach \cite{Derrida91}.
We also show
that the assumption of uncorrelated domain sizes is true at late times, so that
the solution of Nagai and Kawasaki is exact.  Their solution also applies to
systems with long-range interactions, in the limit $\sigma \rightarrow \infty$.

When a domain annihilates, the two adjacent domains combine into one large
domain which is the size of the original three domains put together. For
systems
with short-range interactions, or for long-range interactions with
$\sigma \rightarrow \infty$,
the shortest domain collapses instantaneously with respect to
longer domains (see Figure \ref{FIG:COLLAPSE}). For short-range models,
size differences between nearby domains are of the order of
the growing length-scale $L(t)$ and shorter domains collapse exponentially
faster than any larger ones, with an exponent equal to the size difference
divided by the fixed core-scale $\xi$.  For long-range models,
the smallest domain annihilates instantaneously relative to longer domains
for $\sigma \rightarrow \infty$ due to the pairwise force (\ref{EQN:FORCE}).

Consider a random initial set of domain lengths.
The collapse of the shortest domain corresponds
to taking the smallest length from the set, adding it to
the lengths of its neighboring domains, and replacing the
three lengths by one length equal to their sum.
As this procedure is iterated, the distribution of domain sizes, scaled
by the size of the smallest domain, $\tilde{L}(t)$,
will approach a fixed point distribution.
There will be no correlations formed between
the lengths of different domains because the collapse of the shortest
domain forms a single domain with its neighbors independently
of the other domains in the system.  This is similar, in that respect, to the
``paste-all'' model of Derrida {\em et al.} \cite{Derrida91}.  Since all
higher-point correlations are zero,  the domain size distribution
function describes the fixed point.

If $n_t(l) \delta l$ is the number of
domains of sizes in the interval $[l,l+\delta l ]$
in the entire system, then
\begin{equation}
\label{EQN:DEFN}
N(t) = \int_{\tilde{L}}^{\infty} n_t(l) dl,
\end{equation}
\begin{equation}
\label{EQN:DEFL}
L_\infty = \int_{\tilde{L}}^{\infty} l n_t(l) dl,
\end{equation}
\begin{equation}
\label{EQN:DEFF}
f_t(l) = n_t(l)/N(t),
\end{equation}
\begin{equation}
\label{EQN:DEFSF}
F(l/\tilde{L},t) = \tilde{L} f_t(l),
\end{equation}
where $N(t)$ is the number of domains in the system, $L_\infty$ is the
length of the system, $f_t(l)$ is the probability density of domain
sizes, and $F(x,t)$ is the scaled probability density with respect to the
minimum domain size $\tilde{L}$.  Note that
$n_t(l)$ is the same as used in section \ref{SEC:ENSCALE}. We will
determine the time-independent fixed point distribution $F(x)$.

Each domain that annihilates
combines with the two adjacent domains, each of length $l$
with probability density $f_t(l)$.
Conversely, for each domain that annihilates, one large domain is formed at
length $l$ with a probability density coming from two domains
combining, one of length $l'$ and the other of length $l-l'-\tilde{L}$,
both restricted to be larger than $\tilde{L}$.
In a time interval $\delta t$,  the minimum length will shift by
$\delta \tilde{L}=\tilde{L}(t+\delta t)-\tilde{L}(t)$, and
$n_t(\tilde{L}) \delta \tilde{L}$ domains will annihilate.
This leads to an evolution equation for the number distribution $n_t(l)$:
\begin{equation}
\label{EQN:NUMDIST}
n_{t+\delta t}(l) = n_t(l) + n_t(\tilde{L}) \delta \tilde{L}
  \left[ - 2 f_t(l) + \int_{\tilde{L}}^\infty dl'
  f_t(l') f_t(l-l'-\tilde{L}) \Theta(l-l'-2\tilde{L}) \right].
\end{equation}
The evolution equation for the scaled probability density $F(x,t)$ follows:
\begin{equation}
\label{EQN:SCALEEQN}
\tilde{L} dF(x,t)/d \tilde{L} =
         F(x,t) + x dF(x,t)/dx + F(1,t) \int_1^\infty dx'
        F(x',t)F(x-x'-1,t) \Theta(x-x'-2).
\end{equation}
Using the Laplace transform, $\phi(p,t) = \int_1^\infty dx e^{-p x}F(x,t)$,
we obtain
\begin{equation}
\label{EQN:APPROACH}
\tilde{L} d\phi/d \tilde{L} = -p d\phi/dp+F(1,t) e^{-p} \left[ \phi^2-1
\right].
\end{equation}
For the stationary solution we set $d\phi/d \tilde{L}=0$, use $\phi(0)=1$,
and force consistency
on $d\phi/dp|_{p=0}$ to determine $F(1)=1/2$ \cite{nomoment}.
Hence the stationary solution to (\ref{EQN:APPROACH}) is
\begin{equation}
\label{EQN:LAPLACE}
\phi(p) = \tanh{\left[\text{E}_1(p)/2 \right]},
\end{equation}
where $\text{E}_1(p)=\int_p^\infty dx e^{-x}/x$.
This agrees with Nagai and Kawasaki \cite{Nagai86}.
Expanding the $\tanh$ and doing the inverse transform
leads to the fixed point distribution
\begin{equation}
\label{EQN:FIXEDDIST}
F(x)=    \sum_{i=0}^{[(x-1)/2]}{A_{2i+1} \int_1^\infty \prod_{j=1}^{2i+1}
 \frac{dx_j}{x_j} \delta (x- \sum x_j)}, \ \ \ \ \ \ \ \ x \geq 1,
\end{equation}
where the first sum is for all $i$ such that $x \geq 2i+1$, and
${2^n A_n}$ are the Taylor expansion coefficients of $\tanh(x)$.
The fixed point distribution is piecewise analytic, with
discontinuities at $x=2i+1$ in the $2i$-th derivative for all integer
$i \geq 0$.  The fixed point distribution is simple for small $x$:
\begin{equation}
        F(x)= \left\{ \begin{array}{c}
                0, \ \ \ \ \ \ \ \ \ \ \ \ x \in [0,1),    \\
                1/(2x), \ \ \ \ \ \ \ x \in [1,3], \\
                \end{array} \right.
\end{equation}
but gets progressively more complicated in every succeeding
interval of width $2$.

At finite times, for
systems with short-range interactions, there will be small correlations
between the sizes of different domains because the exponential
suppression of the motion of domain walls relative to those of the
smallest domain is not complete. However, the ratio $R$ of the
average force on a domain wall to the force on
the walls of the shortest domain vanishes as $L(t) \rightarrow \infty$, and so
any correlations will vanish in the scaling, or late time, limit.
For short-range forces,
\begin{eqnarray}
 R &\simeq & \int_1^\infty dx F(x) e^{-\tilde{L}x/\xi} / e^{-\tilde{L}/\xi},
         \nonumber \\
   & \simeq& e^{\tilde{L}/\xi} \int_1^\infty \frac{dx}{2x} e^{-\tilde{L}x/\xi},
         \nonumber \\
  & \simeq & \xi/(2\tilde{L}),
\end{eqnarray}
where $F(x)$ is the fixed point distribution of the scaled domain size
(\ref{EQN:FIXEDDIST}).  Hence $R$ vanishes at late times
as $L \rightarrow \infty$,
and the fixed point distribution will hold in that limit.
For long-range forces, the same calculation leads to
\begin{eqnarray}
 R &\simeq & \int_1^\infty dx F(x) (\tilde{L}x)^{-\sigma} /
\tilde{L}^{-\sigma}, \nonumber \\
   & \simeq&  \int_1^\infty \frac{dx}{2x} x^{-\sigma}, \nonumber \\
  & = & 1/(2\sigma).
\end{eqnarray}
In the limit $\sigma \rightarrow \infty$
this ratio vanishes and the fixed point distribution is valid at all times.

The average domain size is given by
\begin{eqnarray}
\label{EQN:RATIO}
 L(t) &=& \tilde{L} \int_1^\infty x F(x) dx, \nonumber \\
      &=& -\tilde{L} d\phi/dp|_{p=0}, \nonumber \\
      &=& \tilde{L} 2 e^{\gamma_E},
\end{eqnarray}
where we have used the asymptotic form of (\ref{EQN:LAPLACE}),
and $\gamma_E \simeq 0.577$ is the Euler constant. This agrees
with Nagai and Kawasaki \cite{Nagai86} and numerical results \cite{Nagai83}.

The scaled distribution of domain sizes, $F_L(x)= L n_t (x L) /N$, is plotted
against the scaled domain size, $x=l/L(t)$, for $\sigma=0.5$, $1.0$, and
$1.5$ in Figure \ref{FIG:DOMSIZE}.
The exact result for $\sigma \rightarrow \infty$
is also shown.  As $\sigma$ increases, the distribution approaches
the $\sigma \rightarrow \infty$ fixed point distribution.

The growth law of the average domain size, $L(t)=L_\infty/N(t)$,
is derived by matching the flux of annihilating domains $j(\xi)$
to the flux of domains coming from the
minimal domain size $\tilde{L}$. From the discussion leading to
equation (\ref{EQN:COLLAPSE}), the flux of annihilating domains
is given by $j(\xi) = \dot{N}/2 = -\dot{L} N /(2 L)$.
For systems with short-range interactions, the flux at $\tilde{L}$ is given by
$j(\tilde{L}) = - (N/ \tilde{L}) e^{-\tilde{L}/\xi}$, where
$n_t(\tilde{L})= (N / \tilde{L}) F(1)$ is the
number density of domains at scale $\tilde{L}$, they are collapsing
at a rate $-2 e^{- \tilde{L} /\xi}$ (twice the speed of each end), and
$F(1)=1/2$. Equating these fluxes give a logarithmic growth law
\begin{equation}
\label{EQN:SRVEL}
\tilde{L} = \xi \ln{(t/t_0)},
\end{equation}
where $t_0$ depends on the form of the potential $V(\phi)$.

For systems with long-range interactions, with $\sigma \rightarrow \infty$,
we can also obtain the exact growth law.
The flux of annihilating domains is again given by
$j(\xi) = -\dot{L} N / (2 L)$.
The flux at $\tilde{L}$ is now given by
$j(\tilde{L}) = - N / \tilde{L} \tilde{L}^{\sigma}$,
where the domains of scale $\tilde{L}$ are collapsing
at a rate $-2 \tilde{L}^{-\sigma}$.
Equating these fluxes gives $d\tilde{L}/dt = 2 \tilde{L}^{-\sigma}$,
and so $L(t)= 2 e^{\gamma_E} t^{1/(1+\sigma)}$ in the
$\sigma \rightarrow \infty$ limit.  This determines the scaling
amplitude in that limit:
\begin{equation}
\label{EQN:PHI}
        \Phi^\ast(\infty)=g_R^\ast(\infty) =e^{-\gamma_E}/2 \simeq 0.28,
\end{equation}
where the fixed point amplitude used by Lee and Cardy is
$g_R^\ast(\sigma)=\Phi^\ast(\sigma) \left[ 2(1+\sigma) \right]^{1/(1+\sigma)}$
 \cite{Lee93}. This exact result is comparable to the perturbative calculation
by Lee and Cardy who found $g_R^\ast(\infty) \approx 0.33$.

The structure function is calculated by
Kawasaki {\em et al.} \cite{Kawasaki88}
from the Fourier transform of (\ref{EQN:2CORR})
and the domain size distribution function:
\begin{equation}
\label{EQN:FIXSTRUCT}
S(k) = \frac{4}{k^2 L} \frac{1-\left|\tanh{\left[\text{E}_1(ik \tilde{L})/2
        \right]}\right|^2}
        {\left|1+\tanh{\left[ \text{E}_1(ik \tilde{L})/2 \right]} \right|^2},
\end{equation}
which, because $L/\tilde{L}$ is constant,
satisfies the scaling form (\ref{EQN:STRUCT}).

Since domains annihilate independently and only adjacent domain walls
have correlated velocities, the real-space time-derivative
correlation function (\ref{EQN:2TCORR}) is simplified
\begin{eqnarray}
\label{EQN:SRTTCORR}
 T_{\text{SR}}(r,t)
        &=& (\dot{L}/L)^2 \left\{ 4 \left<v^2 \right> \dot{L}^{-2} \delta(x)+
                4 N^{-1} \sum_{<i j>} (-1)^{i+j}
                v_i v_j \dot{L}^{-2} \delta(|x|-|x_{ij}|) \right\},
        \nonumber \\
        &\simeq&
        (\dot{L}/L)^2 \left\{ 4 \left<v^2 \right> \dot{L}^{-2} \delta(x)+
                4 F_L(|x|) \left[ v(|x| L)/ \dot{L} \right]^2 \right\},
\end{eqnarray}
where $x= r/L$, $F_L(x)$ is the domain distribution function scaled with
respect to the average domain size $L$, and
$\left< v^2 \right>$ is the average square velocity of a domain wall.
The equation becomes exact, though the scaling behavior is unchanged,
if we use the RMS velocity of a wall of a domain of
size $l$ in the second line, $v_{\text{RMS}} = |v(l)-\left< v \right>|$,
rather than $v(l)$. For both short-range and $\sigma \rightarrow
\infty$ systems, the average square-velocity is dominated by the
tiny flux of annihilating domains.  Following equation (\ref{EQN:VELSQUARE}),
the average square-velocity will be
\begin{equation}
\label{EQN:SRV2}
\left< v^2 \right>  \sim  \dot{L} L^{-1} \int_\xi^L dl\, v(l) .
\end{equation}
The integral diverges at small $l$ for
both short-range systems and long-range systems with $\sigma >1$,
and the square-velocity is dominated by small annihilating domains,
$\left< v^2 \right> \sim \dot{L}/L$. Hence, the scaling (\ref{EQN:TTSCALE}) of
the time-derivative correlations (\ref{EQN:SRTTCORR}) is broken at $x=0$.
The scaling for $x>0$ holds if $v(x L)/\dot{L}$ is time-independent.
For long-range interactions, with $v(l) \sim l^{-\sigma}$ and $\dot{L} \sim
L^{-\sigma}$, this is true and the $T(r,t)$ scales for $r/L >0$. For
short-range interactions, $v(l) \sim e^{-l/xi}$. Since
$\tilde{L} = \xi \ln{(t/t_0)}$ (\ref{EQN:SRVEL}), scaling only
holds at $x=\tilde{L}/L \simeq 0.28$, and is broken at all other $x$.

\section{Conclusions}
\label{SEC:CONC}

{}From the energetics of scaling 1D scalar systems,
we predict growth laws $L(t) \sim t^{1/(1+\sigma)}$ for all $\sigma >0$.
This growth law is of the same form as
the length-scale given by a collapsing isolated domain (\ref{EQN:FORCE}).
The intuitive picture is of collapsing domains leaving ``voids''
which set the growing length-scale.
These growth laws agree with previous predictions
for $0 < \sigma <1$ \cite{Bray93,Bray94}.
We confirm our predicted growth laws,
with simulations for $\sigma=0.5$, $1.0$, and $1.5$ ---
one value in each of the three regimes of the equilibrium system.
Different boundary conditions and system sizes only affect the late time,
small number regime of our simulations, and so
the thermodynamic limit is well behaved.
The anomalous growth law seen by Lee and Cardy \cite{Lee93} for
$\sigma=0.5$ was the result of large finite-size effects ---  a scaling regime
appears for larger systems than they considered.
The non-analyticities seen by Lee and Cardy \cite{Lee93} for
$0< \sigma \leq 1$ must be controlled in an
infinite system with a fixed initial density.
We find that the equal-time correlations scale for $\sigma=0.5$, $1.0$, and
$1.5$, as do the time-derivative
correlations for $\sigma=0.5$.
For $\sigma \geq 1$ the real-space time-derivative
correlations scales properly for $r > \xi$, however scaling is technically
broken at core-scales due to small rapidly annihilating domains.
For $\xi < r << L$, the correlations can be well approximated by assuming
that domains collapse independently. The approximation becomes better with
increasing $\sigma$.

For purely short-range interactions, and for long-range interactions
in the $\sigma \rightarrow \infty$
limit, we show that the solution by Nagai and Kawasaki
\cite{Nagai86} of the domain-size distribution, which assumes
uncorrelated domain sizes, is exact.  Hence their
expression for the structure factor (\ref{EQN:FIXSTRUCT}), which
satisfies scaling, is also exact. We determine the time-derivative
correlation function, which {\em breaks} scaling at core scales,
and more generally for systems with purely short-range interactions.
For systems with long-range interactions
described by (\ref{EQN:FORCE}) and (\ref{EQN:UPDATE}),
we determine the exact growth law amplitude
in the $\sigma \rightarrow \infty$ limit,
$L(t) = t^{1/(1+\sigma)}/g_R^\ast(\infty) $ with
$g_R^\ast(\infty) = e^{-\gamma_E} /2 \simeq 0.28$.

\acknowledgments

We thank T. Blum, J. Cardy, and B. P. Lee for discussions.

\appendix

\section{Resumming domain-wall interactions in a periodic system}
\label{APP:FORCELAW}

With long-range forces in a periodic system, each domain wall interacts with
an infinite number of images of each other domain wall. Using the interaction
(\ref{EQN:FORCE}), the total force between two domain walls is given by
\begin{eqnarray}
\label{EQN:FORCELAW2}
        f(l) &=& - l^{-\sigma} + \sum_{n=1}^{\infty} \left[
        (n L_\infty-l)^{-\sigma}-(n L_\infty+l)^{-\sigma} \right], \nonumber \\
        &=& -l^{-\sigma}+ \sum_{n=1}^\infty [\Gamma(\sigma)]^{-1}
                \int_0^\infty dt t^{\sigma-1}
                \left[ e^{-(nL_\infty-l)t}- e^{-(n L_\infty+l)t} \right],
\nonumber \\
        &=& -l^{-\sigma}+ \frac{L_\infty^{-\sigma}}{\Gamma(\sigma)}
                \int_0^\infty dx x^{\sigma-1}
                \frac{\sinh{\left[l/(x L_\infty )\right]}}
                {\sinh{\left[ x/2 \right]}} e^{-x/2},
\end{eqnarray}
where $l$ is the closest distance between the domain walls around the loop
 and $L_\infty$ is the system size. The force is attractive along
their nearest separation if the domain walls are of opposite
sign and repulsive if of the same sign.
The interaction energy can be similarly calculated using (\ref{EQN:WALLINT}):
\begin{equation}
\label{EQN:ENLAW}
E(l)-E(L_\infty)= l^{1-\sigma}+\frac{L_\infty^{1-\sigma}}{\Gamma(\sigma-1)}
                \int_0^\infty dx x^{\sigma-2} \left[
                \frac{\cosh(lx/L_\infty) -1}{\sinh(x/2)} \right] e^{-x/2},
               \ \ \ \ \ \ \ \ \ \ \    \sigma \neq 1,
\end{equation}
where we have adsorbed a factor of $(\sigma-1)^{-1}$. An overall
$-1$ factor is introduced between domain walls of opposite sign.
For $\sigma \neq 1$ the integrals are done numerically
and stored in lookup tables for use in the simulations.
For $\sigma=1$, the integrals can be done exactly:
\begin{equation}
\label{EQN:FORCE100}
f_{\sigma=1}(l) = -\pi L_\infty^{-1} / \tan{(\pi l/L_\infty)},
\end{equation}
\begin{equation}
\label{EQN:EN100}
        E_{\sigma=1}(l)-E_{\sigma=1}(L_\infty) =
         \ln{l}+\ln{\left[
        \frac{\sin(\pi l/L_\infty)}{\pi l/L_\infty}\right]}.
\end{equation}

\newpage

\begin{figure}[h]
\caption{A schematic representation of the
1D scalar system with domains of $\phi = \pm 1$ shown.
The domain walls at $r_i$ and $r_j$ are ``positive'',
and the sign of domain walls
alternate along the system.  Domain walls of the same sign
repel each other, while those of opposite sign attract each other.
The dynamics for long-range interactions are described by these
sharp Ising-like walls, as is discussed in text. }
\label{FIG:WALLINT}
\end{figure}

\begin{figure}[h]
\caption{The bottom plots show simulation results of
the scaled average domain-wall velocity
$\left< |v| \right> t^{\sigma/(1+\sigma)}$ vs. time
for $\sigma=0.5$, $1.0$, and $1.5$ with open triangles, squares,
and circles respectively.  The upper plots show
the scaled
square-velocity, $\left< v^2 \right> t^{2\sigma/(1+\sigma)}$,
vs. time with the filled triangles, open squares, and filled circles.
Also shown, for comparison, is the $t^{1/3}$ line expected for the
scaled square-velocity with $\sigma=1.5$.  The systems have periodic boundary
conditions, a {\em fixed} time-step,
and start with 3200, 1600, and 800 domain walls for $\sigma=0.5$,
$1.0$, and $1.5$ respectively.  Note that the statistical errors shown
are not independent.
}
\label{FIG:VEL}
\end{figure}

\begin{figure}[h]
\caption{$\Phi$, the ratio of the measured to expected density
(\protect\ref{EQN:DENSITYSCALING}), plotted
vs. the number of domain walls ($L_\infty/L(t)$) on the left and vs.
the ratio of measured to initial density ($\rho(t)/\rho(0)$) on the right.
The data are for systems starting with 200, 800, 3200, and
12800 domain walls for open triangles, squares, circles, and filled triangles,
respectively. All of the systems have free boundary conditions and
an initial random placement of domain walls.
}
\label{FIG:ALLGROWTH}
\end{figure}


\begin{figure}[h]
\caption{The ratio of the measured to expected density up to a constant
factor (\protect\ref{EQN:DENSITYSCALING}) plotted
vs. the number of domain walls for $\sigma=0.5$.
The top figure corresponds to periodic boundary conditions
with no cutoff to the force law, while the bottom figure
corresponds to free boundary conditions.  The point types are the same
as for Figure \protect\ref{FIG:ALLGROWTH}, with stars corresponding to
systems starting with 1600 domain walls.}
\label{FIG:BOUNDGROWTH}
\end{figure}

\begin{figure}[h]
\caption{The ratio of the measured to expected energy density
(\protect\ref{EQN:DENSITYENERGY}), up to a constant factor, plotted
vs. the ratio of measured to initial density of domain walls. The top figure
is for $\sigma=0.5$, while the bottom is for $\sigma=1.0$. The boundary
conditions are periodic. The point-types are the same
as the previous two figures.}
\label{FIG:ENERGY}
\end{figure}

\begin{figure}[h]
\caption{The distribution of domain sizes, $F_L(x)=L n_t(x L)/N$, as a function
of scaled domain size, $x=l/L(t)$. Note that we have scaled with respect
to the {\em average} domain size, $L(t)$. The triangles, squares, and circles
correspond to $\sigma=0.5$, $1.0$, and $1.5$, respectively. The filled
points correspond to simulations with four times as many updates per factor
of two in time. The line
in the upper right is the {\em exact} result for $\sigma \rightarrow \infty$.
The straight lines indicate power-law fits for small $x$ (see equation
(\ref{EQN:COLLAPSE})), as expected for a uniform rate of domain annihilation.}
\label{FIG:DOMSIZE}
\end{figure}

\begin{figure}[h]
\caption{Real space correlations plotted against $x=r/L(t)$. The top three
figures are the scaled equal-time correlations $f''(x)=L^2 C''(r,t)$
for $\sigma=0.5$, $1.0$, and
$1.5$, while the bottom three figures are the corresponding
scaled time-derivative correlations $g(x)=(L/\dot{L})^2 T(r,t)$.
The open circles correspond to correlations at times greater by a factor
of 4 compares to triangles.   The solid lines are described in the text: the
upper curve represents a parameterless estimate of the correlations, while the
offset straight line represents the expected asymptotic power-law for small
$x$.
The filled circles correspond to simulations
with a finer timestep. The error bars have been suppressed for clarity.
No data is shown for $g(x)$ with $x>1/2$.
}
\label{FIG:ALLCORR}
\end{figure}

\begin{figure}[h]
\caption{For systems with short-range interactions, or with long-range
interactions where $\sigma \rightarrow \infty$, the smallest domain
will annihilate instantaneously with respect to larger domains.
A single large domain is formed of size equal to the smallest
domain and its two neighbours. The filled circles indicate domain walls.}
\label{FIG:COLLAPSE}
\end{figure}

\begin{table}
\caption{The scaling amplitudes of the growth law
(see \protect\ref{EQN:DENSITYSCALING}), where  $g_R^\ast(\sigma) =
\Phi^\ast(\sigma) \left[ 2(1+\sigma) \right]^{1/(1+\sigma)}$
\protect\cite{Lee93}. These amplitudes are for the dynamics specified
in equations (\ref{EQN:FORCE}) and (\ref{EQN:UPDATE}).
\label{TAB:G}}
\begin{tabular}{rcl} \hline
 $\sigma$ &  $\Phi^\ast(\sigma)$ & $g_R^\ast(\sigma)$ \\ \hline
 $0.5$  &   $0.200 \pm 0.005$ & $0.42 \pm 0.01$ \\
 $0.75$ &   $0.180 \pm 0.005$ & $0.37 \pm 0.01$ \\
 $1.0$  &   $0.175 \pm 0.005$ & $0.35 \pm 0.01$ \\
 $1.5$  &   $0.170 \pm 0.005$ & $0.32 \pm 0.01$ \\
 $\infty$ & $e^{-\gamma_E}/2 \simeq 0.28$ & $e^{-\gamma_E}/2 \simeq 0.28$ \\
\hline
\end{tabular}
\end{table}

\begin{table}
\caption{The coefficient of the velocity of fast-moving domain walls,
$v(l) \equiv \beta l^{-\sigma}$ for $l \ll L$,
obtained by fitting figure \ref{FIG:DOMSIZE} by the distribution in
equation (\ref{EQN:VEL}). For $\sigma \rightarrow \infty$, the domains
are uncorrelated and $\beta=1$ (see section
\ref{SEC:SHORTRANGE}).
\label{TAB:BETA}}
\begin{tabular}{rl} \hline
 $\sigma$ &  $\beta(\sigma)$\\ \hline
 $0.5$  &   $0.75 \pm 0.05$ \\
 $1.0$  &   $0.90 \pm 0.05$ \\
 $1.5$  &   $0.95 \pm 0.05$ \\
 $\infty$ & $1$ \\
\hline
\end{tabular}
\end{table}


\begin{references}
\bibitem{soft} Exotic potentials with non-quadratic or asymmetric
minima will lead
to different phase-ordering behavior in 1D systems with short-range
interactions since the growth is not curvature driven.
\bibitem{diss} We assume a velocity-independent
wall mobility, so that velocities are proportional to the force. This
is explicitly shown for $0<\sigma\leq 1$, at low temperatures
with no domain creation, by Lee and Cardy \protect\cite{Lee93}.
\bibitem{size} Our simulations are done with systems of a fixed initial density
and various lengths.  This proved most convenient for choosing the timescales
of simulations. Different initial densities can be formally mapped to a fixed
density with a change of timescale and system size without changing the
scaling functions.
\bibitem{timestep} The distribution of
domains with sizes $l \lesssim l_{\text{min}}$ is changed (made more uniform)
by a geometrically increasing timestep.  This affects
short-distance correlations.
\bibitem{free} Free boundary conditions for the domains, which
are used, have half-strength domain walls at the end of the system and
net ``charge neutrality'' of the signed domain walls.
\bibitem{nomoment} We ignore cases where the average domain size $L(t)$
is infinite. From the small $p$ asymptotics of the
solution, $\phi \simeq 1-O(p^{2 F(1)})$,
these cases correspond to $F(1) < 1/2$ so that
$-d\phi/dp|_{p=0}$, which gives the first moment of $F(x)$, is
infinite. Cases with $F(1) >1/2$ are unphysical --- they
correspond to a vanishing average domain size (see equation \ref{EQN:RATIO}).
\end{references}
\end{document}